# The AI Penalty: People Reduce Compensation for Workers Who Use AI

Jin Kim *, Shane Schweitzer *, David De Cremer, Christoph Riedl *

D'Amore-McKim School of Business, Northeastern University

* Corresponding authors

## Abstract

We investigate whether and why people might adjust compensation for workers who use AI tools. Across 13 studies ($N$ = 4,956), participants consistently lowered compensation for workers who used AI compared to those who did not. This "AI penalty" is robust across different work scenarios and work tasks, worker statuses, forms and timing of compensation, methods of eliciting compensation, and perceptions of output quality. Moreover, the effect emerges in both hypothetical compensation scenarios as well as real monetary compensation of gig workers. We find that perceived effort and perceived agency—the degree to which an individual serves as the originating source of the core intellectual or creative contribution in a task—explain decisions to reduce compensation for AI-users. However, the penalty is not inevitable. Workers who strategically retain creative agency over core tasks recover most of the AI penalty, and employment contracts that make compensation reductions impermissible provide structural means of reducing the AI penalty.

**The AI Penalty: People Reduce Compensation for Workers Who Use AI**

## Introduction

Artificial Intelligence (AI)—computers acting, deciding, and advising in ways that seem intelligent—is expected to increase workers' productivity and transform the way they work [1]. For example, AI helped software engineers at Google write code faster [2], enabled taxi drivers to find customers more efficiently [3], and allowed artists to produce more artworks as well as receive more favorable evaluations over time [4]. Not surprisingly, workers are increasingly using AI for their work. For example, the share of workers reporting AI use at work rose from 16% in 2024 to 21% in 2025, according to a Pew Research Center survey [5].

While AI promises to boost worker productivity and organizational efficiency, it also has the potential to negatively impact the perceived value of workers' contributions [6–10]. Specifically, workers' use of AI may influence how their work is evaluated by others, potentially leading to perceptions that they deserve less credit and therefore lower compensation than workers who do not use AI. Indirectly supporting this notion, recent surveys show that employees are hesitant to admit to their managers that they used AI for common workplace tasks [11,12]. Similarly, people who used AI in the workplace anticipated and indeed received negative evaluations regarding their competence and motivation [9], and people attribute less responsibility to individuals who follow AI advice [13]. These findings suggest that observers and organizations may view workers' AI use negatively—even as AI adoption is actively encouraged in many settings [14]—and this perception can lead to financial penalties for workers. We explore whether people adjust compensation for workers who use AI, identify the perceptual mechanism through which AI use translates into reduced compensation, and show that workers can strategically mitigate this penalty by retaining creative agency. (In this paper, we use the term *compensation* to refer to a decision-maker's monetary valuation of a worker's contribution for a focal task—operationalized as a one-time payment or bonus allocation for that task—rather than a worker's full compensation package, such as salary, benefits, or long-run raises.)

Theoretical frameworks in economics identify countervailing forces through which AI may either reduce or boost worker compensation—for instance, by displacing workers from existing tasks or by raising productivity and creating new ones [15]. Consistent with this ambiguity, empirical evidence is mixed: Some studies find that AI exposure is associated with wage growth [16–19], while others document significant earnings declines, particularly among freelancers and workers with fewer structural protections [20,21]. Yet this evidence is largely observational and considers broad, economic aspects. Compensation decisions, however, are often made by individual managers, clients, and evaluators, whose judgments may be shaped by psychological reactions to AI that operate independently of these broader economic dynamics [22,23]. To address this gap, we draw on equity theory to develop a psychological account of how AI use affects individual compensation decisions.

## Equity Theory and AI Use

Prior research has documented a range of negative psychological reactions to AI, including reduced trust and reluctance to rely on algorithmic outputs [24]. However, the AI penalty we document operates through a different mechanism: Evaluators do not discount the quality of AI-assisted work but rather devalue the worker's perceived contribution to it—shifting the locus of bias from the product to the producer. To explain this, we draw on equity theory [25]. Lay beliefs about how AI participates in work tasks may affect compensation decisions by skewing perceptions of equity [25], thereby making human workers seem less deserving of credit. Observers attend to a worker's input, that is, how much a person contributes to work (e.g., effort, expertise, and other resources), relative to their output, that is, how much the worker receives from the work (e.g., compensation). When a worker's input on a task matches the output, people tend to believe that worker deserves credit for that output; however, when the perceived balance between inputs and outputs is disrupted, such as when inputs are perceived to be less than outputs, people perceive inequity. People attempt to restore equity by adjusting inputs or outputs; decision-makers overseeing workers may adjust outputs, for example, by compensating less [26].

Equity theory's concept of inputs is broad, encompassing effort, expertise, skill, and other resources a worker contributes. In practice, however, research on equity in compensation has focused predominantly on effort as the primary input. We propose that AI disrupts this simplification by decoupling two dimensions of worker contribution that previously co-occurred: effort and creative agency—the degree to which an individual serves as the originating source of the core intellectual or creative contribution in a task. When effort and authorship move together—as they typically do in non-AI-assisted work—there is little reason to distinguish them. AI makes this distinction consequential by allowing workers to legitimately produce high-quality outputs while varying independently in how much effort they exert and how much creative authorship they retain.

First, the use of AI may reduce how much effort people believe workers put into their output. More so than previous technologies, significant emphasis has been placed on AI's ability to increase efficiency [27]. If a worker produces the same output more efficiently, perceivers may judge that the person worked less hard, that is, they have decreased their input for the same resulting output, thereby disrupting the ratio of input to output compared to the prior status quo. Second, AI systems may be perceived as reducing the agency that a worker exerts in completing tasks. We define creative agency as *the degree to which an individual serves as the originating source of the core intellectual or creative contribution in a task*. Because AI is typically viewed as capable of learning and autonomous behavior [28], if a worker uses AI for a task, perceivers may determine that the AI exerted more authorial control over the output, accordingly splitting the relative input between the worker and the AI, thereby also disrupting the ratio of worker input to their output. Examining relative perceived contributions of workers using AI or not provides a unique lens on issues of using AI for work. Rather than merely perceiving the worker as having different capacities, we study how the *process* of working may be perceived differently as a function of using AI.

We conducted 13 experiments (including six preregistered experiments; $N = 4,956$ individual participants in total) to investigate how workers' use of AI affects people's decisions to compensate those workers. We find robust evidence that people consistently reduce compensation for workers who use AI, a phenomenon we term the "AI penalty." This effect is observed across different tasks, worker statuses, and organizational constraints. We find that this AI penalty operates primarily by shaping evaluators' perception of workers' input—they are perceived as exerting less effort and having lower agency over their work—which reduces how much credit they think they deserve.

Our findings offer valuable insights into how the use of AI can affect worker compensation. By demonstrating that AI use reduces the extent to which workers are perceived to deserve credit for their work, we contribute to the literature on the *psychological* effects of AI in the workplace and labor market [9,29–31]. Moreover, our results highlight possible ways that inequality among workers may be exacerbated through the introduction of AI use, as compensation is more likely to be reduced for workers for whom reduced compensation is considered more permissible (whose baseline compensation is likely to be relatively low to begin with).

## Results

### Evidence for AI Penalty on Compensation

To establish baseline evidence for the effect of AI use on compensation, we conducted two studies (Study 1, $N = 303$; Study 2, $N = 359$). In both studies, participants imagined that they were running a small business for which they hired different graphic designers to create social media ads. Participants were randomly assigned to either learn that the designer intended to use AI for the task or receive no information about the designer's AI use, and then answered how much they would pay the designer (using a slider scale from \$0 to \$100 with \$1 increments). Participants in Study 1 offered a smaller hypothetical payment to the designer who intended to use AI ($M = \$33$, $SD = \$14$) than to the designer whose AI use was not indicated ($M = \$47$, $SD = \$8.5$), $t(301) = -10.70$, $p < 0.001$, $d = -1.23$; Figure 1a). We find consistent results in Study 2 which used different wording ($M_{\text{No AI}} = \$35$ vs. $M_{\text{AI}} = \$47$, $p < 0.001$, $d = -1.03$). Together,

Studies 1 and 2 provide initial evidence of the AI penalty: People reduce compensation for workers who use AI (relative to workers who do not use AI).

To further substantiate the AI penalty, we tested for its presence with *real* compensation for *real* gig workers (Study 10, $N = 230$). To do so, we first recruited participants (workers, $n = 60$) to write social media posts either using AI or not. We then recruited a second group of participants (managers, $n = 140$) to take on the role of manager that allocates real monetary bonuses to the workers who purportedly had used AI or not (see the *Methods* section and *SI* for details). Managers were randomly assigned information indicating whether each worker had used AI or not, but this information was manipulated independently of the worker's actual AI use. Managers gave smaller bonuses to workers who purportedly had used AI ($M$ = \$0.35, $SD$ = \$0.21) than to workers who purportedly had not ($M$ = \$0.65, $SD$ = \$0.21, $t(139) = -8.74$, $p < 0.001$, $d_z = -0.74$, in a dependent $t$-test; Figure 1b). We thus find evidence of the AI penalty with real bonuses to real gig workers. We also conducted regression analyses which held constant workers' actual performance (see *SI*, section 25 and Table S6). To do so, we recruited yet another group of participants (judges; $n = 30$) and asked them to rate the effectiveness of the social media posts (i.e., the likelihood that they would "encourage readers to learn more about the product"); these ratings served as our measure of workers' performance. The analyses revealed results consistent with those reported above: The purported use of AI by workers led managers to significantly reduce compensation for workers ($b = -15.45$, $SE = 1.46$, $t(418) = -10.61$, $p < 0.001$), even when worker performance was held constant. We extend these results with a between-subjects experimental design (Study 11, $N = 505$). As before, participants acted as a manager deciding bonuses of real workers, but this time they were shown only one social media post (see the *Methods* section and *SI*, section 13). We find consistent results of managers giving smaller bonuses to their workers who purportedly had used AI ($M$ = \$2.28, $SD$ = \$1.58) than to their workers who purportedly had not ($M$ = \$3.27, $SD$ = \$1.46), $t(503) = -7.25$, $p < 0.001$; see Figure 1c and *SI*, section 26). These results suggest that managers reduce bonuses for workers who use AI not only when those workers are *juxtaposed* with workers who do not use AI (Study 10), but also when these workers are evaluated in isolation (Study 11). That is, the AI penalty is robust across both *comparative* and *non-comparative* contexts (i.e., whether AI use varies between workers or not) and across both *competitive* and *non-competitive* contexts (e.g., whether workers are competing for a fixed pool of bonuses or not).

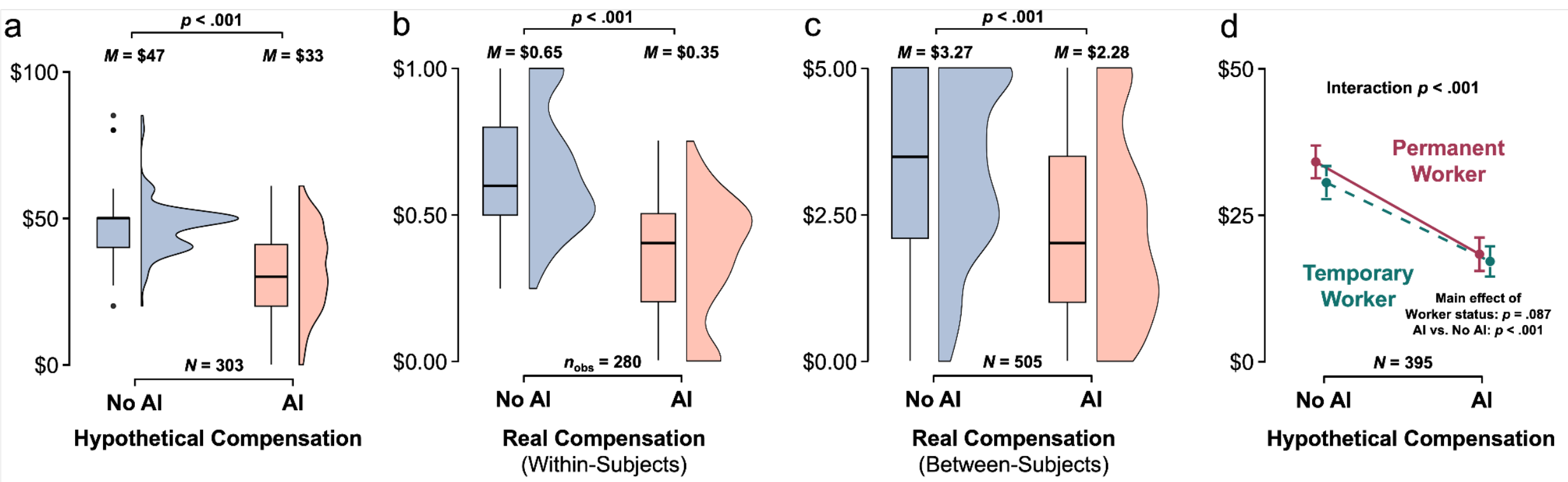

**Fig. 1 | AI Penalty Is Robust Across Contexts and Settings. a** Participants reduced compensation for workers who used AI compared to workers who did not use AI when the workers' compensation was hypothetical (Study 1 among many other studies). **b** Compensation similarly declined for real pay (Study 10). **c** Compensation also declined when decision makers were not directly comparing between workers (Study 11). **d** This AI penalty affects both permanent and temporary workers (Study 3).

## AI Penalty Is Robust Across Contexts and Settings

In Studies 3-6, we examined whether the AI penalty is robust across different contexts and settings. First, we tested whether the AI penalty differed depending on whether the worker was hired permanently (e.g., full-time or salaried) or temporarily (i.e., as a freelancer). It is possible that, due to an ongoing professional relationship with permanent workers, decision-makers may feel more motivated or pressured to retain compensation rates for permanent workers than temporary workers. However, if compensation decisions are driven predominantly by perceived contributions to the work, relational status to the worker may not meaningfully affect those decisions. Building on the same vignette template as before, we assigned study participants to one of four conditions in a 2 (Temporary vs. Permanent Worker) × 2 (AI vs. No AI) between-subjects design (Study 3, $N = 395$). Participants gave a smaller hypothetical bonus to a worker who used AI than one who did not—both when the worker was a freelancer ($M_{\text{AI}}$ = \$15 vs. $M_{\text{No AI}}$ = \$27; $p < 0.001$) and when the worker was a salaried worker ($M_{\text{AI}}$ = \$16 vs. $M_{\text{No AI}}$ = \$30; $p < 0.001$; see Figure 1d). We find no difference in compensation for workers of different status ($p = 0.087$) and no interaction effect between the status of the worker and the use of AI ($p = 0.41$). We find consistent results in a study with a different scenario (Study 4, $N = 398$; see the *Methods* section and *SI*, Section 6). Across both Studies 3 and 4, participants reduced compensation for workers who used AI, and this reduction was similar for temporary and permanent workers.

To test whether a long history of working with the same employer may buffer the AI penalty—for example, if employers are more reluctant to penalize employees they have worked with extensively—we randomly assigned participants to one of four conditions in a 2 (No History vs. Long History) × 2 (AI vs. No AI) between-subjects design (Study 5, $N = 200$).

Participants reduced the compensation of workers who used AI—both when the worker had no history of working with them ($M_{AI}$ = \$36 vs. $M_{No\ AI}$ = \$48; $p < 0.001$) and when they had a long history of working with them ($M_{AI}$ = \$40 vs. $M_{No\ AI}$ = \$53; $p < 0.001$) in simple effects analyses. A two-way ANOVA revealed a nonsignificant interaction between the History condition and AI condition ($p = 0.86$), but significant main effects of both the History condition and the AI condition ($p$s < 0.004). Neither the worker status nor prior collaboration history protected workers from people's tendency to reduce compensation for workers using AI. These results suggest that the AI penalty is robust across workers with different statuses and that it may generalize to a wide range of employment situations.

To test whether productivity gains reduce the AI penalty, we examined compensation decisions when AI use is framed as increasing worker productivity (Study 6, $N = 401$). Because we described workers' productivity growth, most participants (66%) reported an intent to increase compensation for the workers. More importantly, however, we still find that the use of AI had a negative effect on compensation: The percentage of participants intending to increase the compensation for workers in the No AI condition was 79%, but this percentage significantly decreased to 52% in the AI condition, $\chi^2$ (1) = 30.71, $p < 0.001$. This decrease in the proportion (of participants intending to increase the compensation) was comparable for part-time and full-time employees (see the results on the interaction term [$p = 0.13$] in the logistic regression analysis in *SI*, Section 21). Thus, while recognition of workers' productivity growth led people to increase compensation for them, the workers' use of AI still had a negative effect on their hypothetical compensation, counteracting the positive effect.

**The AI Penalty Is Not a General External-Help Penalty**

A reasonable alternative account of the results discussed so far is that any form of external assistance (and not necessarily *AI* assistance) should reduce compensation for a worker, because the worker is less the driver of the work output. Under this account, the "AI" penalty would not be unique to AI at all, but rather workers would be penalized merely for receiving assistance. We tested this by comparing three conditions (Study 9, $N = 471$): No Assistance, AI Assistance, and Human Assistance (i.e., help from another human). As in our previous studies, AI assistance reduced compensation relative to No Assistance ($d = -0.94$). Critically, however, Human Assistance *increased* compensation relative to No Assistance ($d = 0.35$). In other words, human collaboration, despite introducing external assistance, appears to *add* perceived inputs to the production process rather than diminish the focal worker's contribution. These findings suggest that the AI penalty is not driven by a generic external-attribution mechanism but by something specific to how AI—unlike human collaborators—undermines the perception that the output reflects the worker's own central contribution, which we further explore over the following sections.

**Identifying the Mechanism: Credit Deservingness**

Having established the robustness of the AI penalty across contexts and ruled out a generic assistance penalty, we next investigated *why* evaluators reduce compensation for AI-assisted workers. We theorize that AI use is associated with lower perceived credit for workers' output, and that perceived credit, in turn, is associated with compensation, such that AI use is indirectly associated with lower compensation via perceived credit. We test this hypothesis using the graphic designer scenario while incorporating a three-item measure of credit deservingness (Study 7, $N = 303$). We find the indirect effect of AI use on compensation through credit deservingness is significant and negative ($\beta = -6.33$, 95% CI [−8.81, −4.05], $p < 0.001$; see *SI*, section 22 and Figure S4). Specifically, the worker's use of AI significantly reduced the extent to which participants perceived the worker as deserving credit for the work ($a = -1.56$, $p < 0.001$), and lower perceived credit deservingness in turn predicted lower compensation ($b = 4.04$, $p < 0.001$). These findings provide initial evidence that the AI penalty arises because evaluators attribute less credit to workers who use AI.

Another study (Study 8, $N = 281$) tested for credit deservingness as the mechanism underlying the AI penalty and examined whether the indirect effect through credit deservingness may be attenuated when it is impermissible to act on the AI-related judgments. If perceived credit deservingness indeed drives compensation decisions, then structural constraints that render it impermissible to translate those judgments into financial penalties should attenuate the effect. While evaluators may still privately view AI-assisted workers as deserving less credit, the presence of specific employment contracts creates a boundary where acting on that deservingness assessment becomes procedurally invalid. This account was tested by manipulating whether the worker's bonus was stipulated in an employment contract. We find the indirect effect through credit deservingness is indeed attenuated when reducing compensation was less permissible (index of moderated mediation = 10.06, 95% CI [0.37, 20.49]; see *SI*, section 23). This suggests that formal compensation structures can partially constrain the AI penalty by weakening the link between credit judgments and pay, though the penalty may persist through other psychological channels that institutional arrangements leave unaffected.

**Effort and Agency as Distinct Pathways to the AI Penalty**

To delve deeper into the mechanism underlying the AI penalty, we build on equity theory and test whether the reduced compensation is explained by perceived reduction in worker inputs.

That is, workers who use AI are compensated less, possibly because people think they contributed less. We specifically explore two types of input: "sweat" (effort) and the strategic choice of the worker to retain agency. Going beyond measurement, we experimentally manipulated the worker input (effort and agency) in addition to AI use in a 2 (AI vs. No AI) × 2 (Effort: Low vs. High) × 2 (Agency: Low vs. High) between-subjects design (Study 12, $N$ = 809). Specifically, we experimentally manipulated the strategic choice of the worker to use AI for peripheral vs. core tasks. That is, they either used AI in a way that retained their (creative) agency and ownership in their work, or in a way that diminished it. Furthermore, we manipulated how much effort they exerted (low [30 minutes] vs. high [4 hours]), as well as AI use itself (AI vs. No AI). Participants imagined the same scenario as in Study 2 (running a small business and hiring a graphic designer). We first estimated a saturated structural model including direct paths from all experimental manipulations to the final outcome (Worker Compensation). These direct paths were non-significant ($p > 0.15$ for all), suggesting the effects of the manipulations on the bonus were fully mediated by the proposed mechanisms. We also included interaction terms between the manipulations which were also non-significant on all paths. We therefore trimmed these paths to estimate a more parsimonious over-identified model. This model fit the data well ($\chi^2(3) = 8.06$, $p = 0.045$; CFI = 0.998; RMSEA = 0.046).

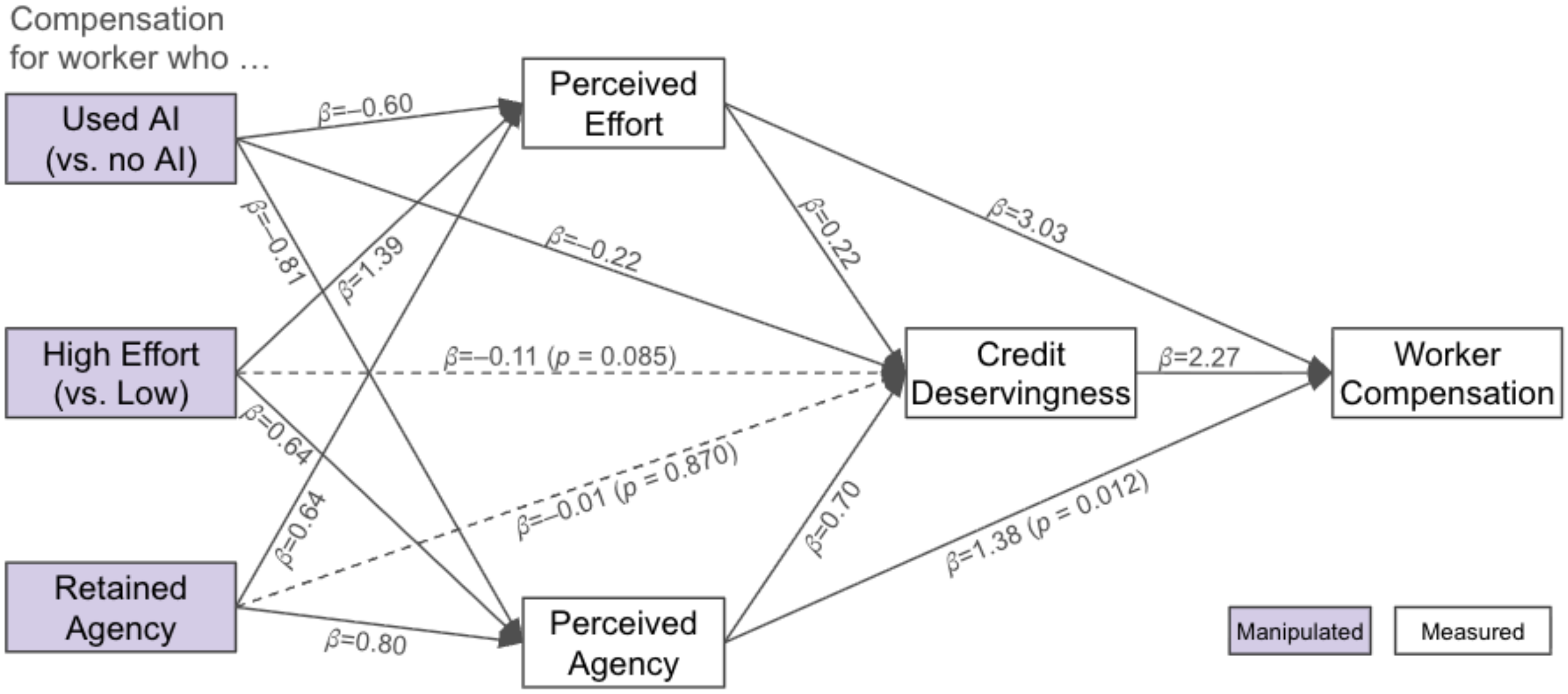

**Fig. 2 | AI use affects compensation through changes in perceived effort, agency, and credit deservingness.** The total effect of AI use on compensation is negative and significant ($\beta = -5.00$, 95% CI [-6.16, -3.79], $p < 0.001$) and operates entirely through the mediating pathways (direct path from AI to bonus is non-significant in the saturated model). Workers who strategically retained agency over core tasks could recover most of this AI penalty (total indirect effect of agency on compensation: $\beta = +4.60$, $p < 0.001$, 95% CI [3.47, 5.76]). Values represent unstandardized coefficients from the over-identified model, which demonstrates excellent fit

($\chi^2(3) = 8.06$, $p = 0.045$; CFI = 0.998; RMSEA = 0.046). Inference is based on 5,000 bootstrap resamples and all paths are significant at $p < 0.001$ unless noted otherwise (Study 12; $N = 809$).

We find strong evidence for a significant AI penalty that operates entirely by changing people's perception of worker inputs and credit deservingness (Figure 2). First, AI use causes observers to perceive significantly less effort ($\beta = -0.60$; $p < 0.001$) and lower agency ($\beta = -0.94$; $p < 0.001$). In line with equity theory, both of these worker inputs have strong effects on worker compensation (effort: $\beta = 3.03$, $p < 0.001$; agency: $\beta = 1.38$, $p = 0.012$). Second, AI use lowers workers' perceived credit deservingness ($\beta = -0.22$, $p < 0.001$), even after controlling for their effort and agency. This is a small, but real, bias against AI. Additionally, there is a serial pathway where perceived effort and agency also affect credit deservingness, which reduces compensation.

The total effect of AI use on compensation through changed perceptions is negative and large ($\beta = -5.00$, 95% CI [-6.16, -3.79], $p < 0.001$). Crucially, the strategic choice in how the worker used AI plays a major role and has a total positive effect on compensation ($\beta = +4.60$, $p < 0.001$, 95% CI [3.47, 5.76]). The indirect effect of retaining agency on compensation recovers approximately 92% of the total AI penalty ($\beta_{\text{agency}} / \beta_{\text{AI}} = 4.60 / 5.00$). Workers who retain agency over the core aspects of their work are seen as working harder ($\beta = 0.64$, $p < 0.001$) in addition to the positive effect on perceived agency itself. Thus, an AI-assisted worker who retains high agency receives compensation nearly statistically indistinguishable from an unassisted worker. This suggests that evaluators make a qualitative judgment about ownership and authorship that carries a distinct signal about deservingness beyond effort. Notably, we find no significant AI × Agency interaction effect (tested separately in a saturated model), suggesting agency does not reduce the credit penalty specifically attached to AI use. Instead, agency operates additively—it raises the baseline perception of effort and credit regardless of whether AI was involved. This suggests that while AI use carries a significant compensation cost, workers who strategically retain authorship over core tasks can substantially mitigate this penalty. While effort can statistically recover the AI penalty as well, it represents an inefficient strategy for workers that would require substantial increase of their work input to offset the AI devaluation, while retaining agency restores pay without necessarily increasing inputs. To substantiate our claim that agency is a new type of input that operates through a distinct pathway, in the *SI*, we show (a) effort and agency are two distinct constructs, and (b) contrast our extended model with a pure equity theory model which we find fits the data significantly worse.

**Beyond Creative Work: the AI Penalty in Analytical Tasks**

The findings of the preceding studies were observed in tasks associated with creative work (e.g., creating ads, webpages, or social media posts) where authorship norms are salient and originality is expected, raising the question of whether agency operates only when creative ownership is at stake. Study 13 ($N = 301$) sought to replicate these findings with a routine task (designed and conducted after Study 12). Participants acting as hypothetical managers were asked to determine the pay for workers who "compile and summarize quarterly sales data into a standardized report." We again manipulated whether the worker used AI and measured perceived effort and agency. Unlike the preceding studies, we did not include a separate measure of credit deservingness; because bonus allocation itself constitutes a consequential act of credit attribution, the direct path from agency to bonus captures credit judgments behaviorally, allowing us to test whether agency drives compensation without the intermediate attitudinal step. We fit a saturated parallel mediation model which naturally had nearly perfect fit (Figure 3). We find that AI use reduces both perceived effort ($\beta = -2.01$, $p < 0.001$) and perceived agency ($\beta = -1.96$, $p < 0.001$) by roughly equal magnitudes. Agency significantly predicts bonus ($\beta = 4.25$, $p < 0.001$), while effort has a substantially smaller and borderline effect ($\beta = 1.96$, $p = 0.103$ in bootstrap inference and $p = 0.040$ with parametric estimates). The indirect path through agency is significant and substantial ($\beta = -9.25$, $p < 0.001$). The indirect path through effort is not significant ($\beta = -2.95$, $p = 0.132$). The direct effect of AI on compensation becomes negligible ($\beta = -0.17$, $p = 0.926$), suggesting the entire AI penalty is almost entirely transmitted through perceived agency. In the *SI*, we also explore a serial mediation model for which we find strong support. This analysis suggests that managers do not directly reward effort but strongly consider effort when forming judgments about the worker's agency; they reward perceived authorship, and effort only influences pay by shaping authorship judgments.

These results provide further support for the notion that agency and effort are two distinct inputs evaluated by managers. While AI use reduces perceived effort and agency by similar amounts, compensation decisions are driven primarily by agency: The indirect effect through agency is roughly three times larger than the indirect effect through effort (-9.25 vs. -2.95), suggesting that agency is a distinct path from effort. This distinction is particularly noteworthy given the two constructs' high correlation ($r = 0.80$, $p < 0.001$): Despite sharing most of their variance, the path via agency is stronger in a setting where collinearity works against finding such differentiation, thus providing conservative evidence for agency's unique role. Substantively, this suggests AI decouples agency and effort by enabling judgments of high agency with low effort (e.g., reviewing and correcting AI output) and, conversely, judgments of high effort with low agency (e.g., manually entering data). Furthermore, these findings suggest that agency also operates for tasks where authorship claims may be less clear than for creative tasks like creating ads, webpages, or social media posts. We find that managers rated the originality of financial report significantly higher when they thought the worker compiled it alone compared to when they used AI (5.48 vs. 3.66, $t = -10.927$, $p < 0.001$) despite the task having no creative component in the way that the aforementioned creative tasks have. This suggests that managers conceive of agency as a higher-order construct rather than a narrow creative one indicating whether the person was the originating source of the work product. However, the underlying reason likely differs: For creative tasks, agency signals authorship; for routine tasks, agency signals accountability and epistemic warrant. In sum, it seems that agency captures the fundamental evaluation of whether a person can be said to *have done the work*—which matters for creative credit, trust, accountability, and desert simultaneously.

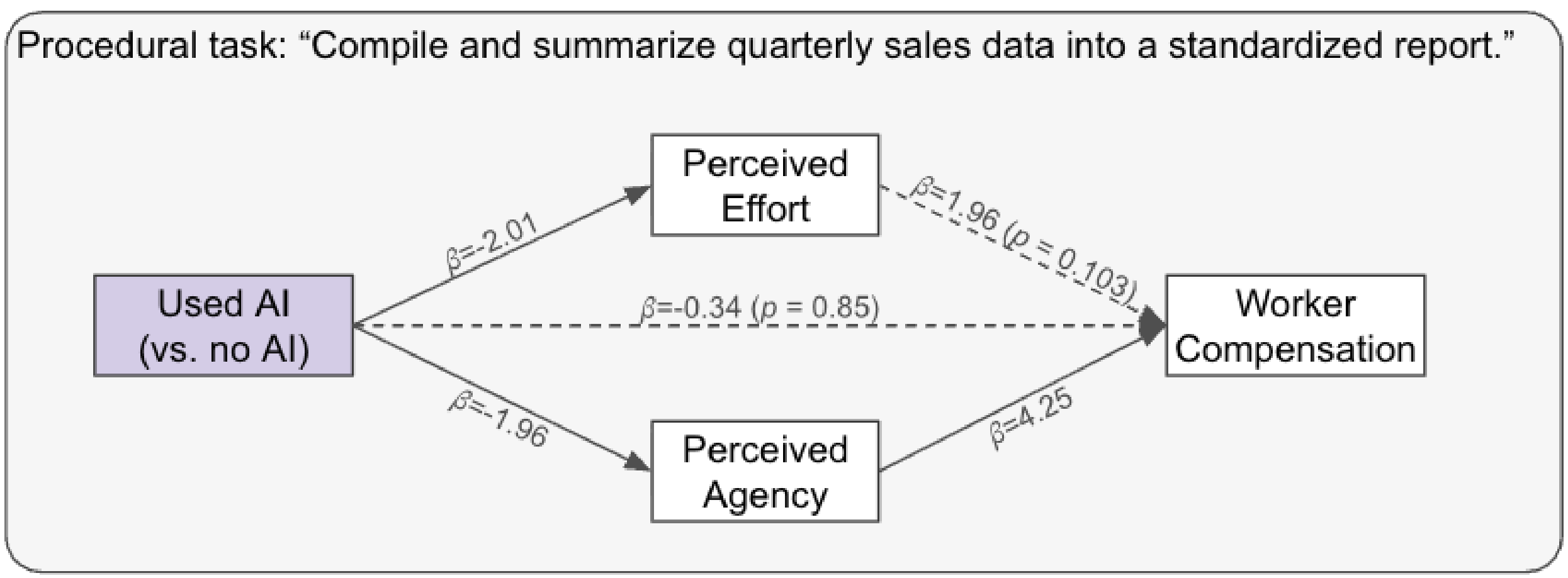

**Fig. 3 | AI use affects compensation through changes in perceived effort and agency in routine tasks.** AI use reduces both perceived effort and agency by roughly equal magnitudes. Agency significantly predicts compensation, whereas the indirect effect of effort is borderline ($p = 0.103$ with bootstrap inference, $p = 0.040$ with model-based inference). Effort and agency fully mediate the effect of AI with no significant direct effect. Values represent unstandardized coefficients from the saturated model. Inference is based on 5,000 bootstrap resamples and all paths are significant at $p < 0.001$ unless noted otherwise (Study 13, $N = 301$).

## General Discussion

Across 13 studies, we find evidence for the AI penalty—people's tendency to reduce compensation for workers who use AI relative to workers who do not. Specifically, we find consistent evidence for the AI penalty in both real monetary compensation decisions involving gig workers and hypothetical scenarios. The effect was specific to AI assistance rather than collaboration with other humans and generalized across diverse work contexts, worker relationships and statuses, compensation structures, payment timing, and elicitation methods. We further uncover a psychological process underlying the AI penalty: that AI use reduces perceived credit deservingness by crowding out two key inputs to compensation judgments—perceived effort and agency over the work.

**Implications of the Findings**

Our results carry important implications for how compensation decisions are made as AI pervades organizational environments, as well as potentially consequential societal ramifications for who stands to be penalized the most. First, we uncover how the ways in which people construe the process of doing work is impacted by the presence of AI. By default, it seems that people assume that when workers use AI, they contribute less authorship to their work output, as well as less effort. These perceptions affect how deserving people think workers are of the compensation they may receive. These perceptions, however, are malleable. Our findings point to two distinct levers for mitigating the AI penalty. At the individual level, workers who strategically retain creative agency recover a substantial portion of the compensation penalty. This effect operates not by changing the work output itself, but by changing the process of how it is produced. At the institutional level, formal compensation structures such as employment contracts attenuate the penalty by constraining evaluators' ability to act on biased deservingness judgments—even when those judgments persist. This could be a crucial mechanism for organizations that want to encourage workers to use AI. Without such structural assurances, calls to adopt AI amount to asking workers to accept a pay cut for being more efficient.

Our results also have important implications for both workers and policymakers. When decision-makers have the flexibility to reduce pay to AI users, they reliably tend to do so, regardless of whether the worker is temporary or permanent; however, if employment contracts are in place, the AI penalty is reduced. Therefore, if permanent and temporary workers who use AI are not meaningfully differentiated in terms of compensation *decisions*, employment contracts may serve as a structural (rather than psychological) mechanism by which temporary workers are penalized for using AI whereas permanent workers are not. For permanent workers, employment contracts with safeguards against reducing pay are generally the default; for temporary workers, the opposite is true. For example, independent contractors and freelancers who operate under general agreements regarding project scopes or hourly rates rather than fixed compensation terms may be particularly vulnerable to compensation reductions when using AI. Similarly, workers under implied or informal contracts, those under flexible compensation schemes (e.g., based on performance), those whose compensation is adjustable during probationary periods, or even those who work in startups or tech companies (where employment agreements might emphasize equity, stock options, or future profits rather than fixed compensations), may also be especially vulnerable to the AI penalty.

Thus, our findings suggest that workers negotiating employment contracts and expecting to use AI in their work may want to be especially mindful of how their compensation will be determined and how they may be vulnerable to the AI penalty. Likewise, policymakers should be aware of the need for increased protections for vulnerable populations of workers. Temporary workers are already more likely to see wage reductions and less likely to receive benefits like healthcare through their employers [32]. Our results indicate a psychological process that may exacerbate this problem. Contracts are just one structural means of mitigating this disproportionate penalization. Policy may need to intervene to offer greater protection for workers in flexible arrangements or incentivize employers to increase transparency in how AI use influences compensation decisions.

Our findings also suggest that there may be something unique about AI that affects people's judgments differently than other tools. Evaluators penalized workers for being assisted by AI but not for being assisted by another human worker. This result is interesting because what separates AI from other tools is arguably the human-likeness of AI. Accordingly, one might think that assistance from AI and assistance from another human might lead to similar reductions in compensation (at least to the extent that AI is like a human). But this is not what we found. If there is indeed something unique about AI (apart from its human-likeness) that drove the AI penalty, it underscores the need for further research to understand exactly what it is about AI that produces phenomena like the AI penalty.

**Limitations and Other Directions for Future Research**

To investigate the effect of AI use in isolation, we ensured in most of our studies that the only difference between the AI and No AI conditions was the disclosure of the workers' use of AI. In other words, apart from the information on whether workers used AI or not, we eliminated all other sources of variation, such as differences in workers' effort or performance. Although this careful manipulation ensured that any effect on worker compensation would be exclusively attributed to AI use (and any resulting inferences), it may have had an unintentional negative impact on the ecological validity of our findings. In real work settings, AI use can impact various factors of work that can affect decisions on worker compensation. That is, our experimental designs do not capture the full complexity of competitive labor markets—where underpaying productive AI-assisted workers risks losing them—which suggests our findings are conservative: Market pressures may attenuate the behavioral penalty, but the underlying equity-based mechanism we identify would persist.

Moving beyond the scope of our investigation, future research may explore additional psychological mechanisms that could produce effects similar to the AI penalty (e.g., workers using AI being penalized in social evaluations relative to those using traditional tools because they are perceived to be lazier [9]). Many of our studies used hypothetical scenarios, though Studies 10 and 11 found the same effect with real monetary compensation for real gig workers. Future research should examine AI-related compensation decisions in more naturalistic workplace settings.

We theorize and test an account of how perceptions of AI's impact on the process of work affects compensation decisions. However, using AI for work tasks also negatively affects perceptions of worker traits and capacities [9]; it also changes how people construe tasks [33], systematically viewing them as less complex. While we specifically manipulate and test distinct mechanisms, and find robust effects across types of tasks, future research should more systematically test and synthesize when and how these distinct perceptions of work with AI impact human workers. We predict that a constellation of perceptions about workers, the work process, the tasks themselves, and AI may jointly shape downstream outcomes for human workers.

## Conclusion

Our research shows that people consistently reduce compensation for workers who use AI relative to workers who do not—a phenomenon we term the "AI Penalty." This effect was observed both for real monetary compensation and hypothetical compensation, and it was robust across various work types, worker statuses, forms and timing of compensation, and methods of eliciting compensation decisions. We identify a psychological mechanism underlying this effect: AI use reduces perceived credit deservingness, and this reduction in credit arises because AI use changes how evaluators perceive key worker inputs. Specifically, AI use lowers perceived effort and—more importantly—perceived agency over the work. Both inputs predict compensation, but agency emerges as the dominant pathway, suggesting that workers perceived as retaining authorship over core tasks can substantially offset the AI penalty, even when AI is used. Moreover, the AI penalty through credit deservingness diminished when reducing compensation was less permissible. Our findings highlight the potential of AI use exacerbating inequality in worker compensation, as workers without contractual protections may be more vulnerable to the AI penalty. As AI reshapes our work and the labor market, it will become increasingly critical to consider and address its potential impacts on workers and shared expectations of economic equity.

## Methods

### Research Transparency Statement

This research was approved by the Institutional Review Board at [redacted for peer review] (IRB #: 23-12-14). Funding was provided by [redacted for peer review]. We have no conflicts of interest to disclose. All participants were recruited from Prolific and were adults who provided informed consent. We did not exclude any participants who completed the study from our analyses. Six of the 13 studies were preregistered (Studies 1, 2, 7, 10, 11, and 13). All preregistrations, data, and analyses code will be available prior to publication on the project's Open Science Framework page: https://osf.io/awhbg/

### Study 1

Study 1 tested whether people reduce hypothetical compensation for workers who use an AI system to complete a task. We recruited 303 workers from Prolific and randomly assigned them to one of two conditions: AI versus No AI. In both conditions, participants read and imagined a scenario in which they were running a small business. Participants imagined that they have hired different graphic designers to create social media ads throughout the year and that they have paid the designers "between $40 and $60 for about an hour's work." They further imagined that they had found a new graphic designer who agreed to create a new ad for them. In the AI condition, the scenario continued with the sentence: "The graphic designer asked if they could use an AI system to assist in creating the ad, and you agreed." This sentence was absent in the No AI condition. Then, in both conditions, participants read that the graphic designer estimated it would take about an hour to create the ad. We then asked participants, "How much payment would you offer them?" Participants determined their hypothetical compensation for the graphic designer on a 101-point slider scale from $0 to $100 with $1 increments (and the slider button initially anchored at $0). After answering this dependent measure, participants provided demographic information (age and gender) to complete the study. The materials of this and all other studies are presented in *SI*. For Study 1 materials, see *SI*, Section 3.

### Study 2

Study 2 was very similar to Study 1. The only notable difference was that Study 2 participants entered the study immediately after completing another study for an unrelated project. That is, Study 2 participants were recruited for two studies (one unrelated study for another project, in addition to Study 2), whereas Study 1 participants were recruited only for Study 1.

The design of Study 2 was identical to that of Study 1, randomly assigning participants to either the AI or No AI condition. Study 2 recruited a slightly larger sample of participants ($N$ =

359), but all other aspects of the study were the same as in Study 1, except for very minor differences (e.g., having a preamble before the scenario for a smoother transition from the unrelated study [“We have one last unrelated question”] or changes in wording such as “Over the course of the year, you have hired different graphic designers…” vs. “Throughout the year, you have hired different graphic designers…”). For Study 2 materials, see *SI*, Section 4.

**Study 3**

Study 3 tested the robustness of findings from Studies 1 and 2 by featuring four main differences. First, Study 3 investigated whether the AI penalty might differ for temporary and permanent workers. That is, whereas Studies 1 and 2 examined people’s decisions to compensate only the workers who were hired temporarily (i.e., graphic designers who were hired for a one-off task of creating a social media ad), Study 3 directly manipulated the type of workers (temporary or permanent workers) to test whether the AI penalty might differ between the different types of workers. Second, Study 3 tested whether the reduction observed in one kind of compensation (i.e., a “payment” that implies a required, agreed-upon price to pay for service rendered) might also be observed for a different kind of compensation (i.e., a “bonus” which is an optional, additional payment for a service). Third, Study 3 tested whether reduction in compensation would occur after the task is completed rather than before the task is completed. Lastly, Study 3 tested whether reduction in compensation would occur even when the quality of the workers’ output is held constant.

Study 3 used a scenario adapted from that of Study 1 and employed a 2 (Temporary vs. Permanent Worker) × 2 (AI vs. No AI) between-subjects design. As in Study 1, participants across the four conditions ($N = 400$) read and imagined a scenario in which they were running a small business. In the Temporary Worker condition, participants imagined that they “hired a graphic designer from an online freelancing platform to create a social media ad for [their] business,” whereas in the Permanent Worker condition, participants imagined that they “assigned a task to [their] salaried graphic designer to create a social media ad for [their] business.” Participants in all four conditions further read that they “received the ad they created… and [were] satisfied with it. The quality of the ad matche[d] that of other ads [they] have recently used for [their] business.” This scenario detail ensured that all participants, regardless of their condition, perceived the worker’s output to be of uniformly high quality. Participants then read about a reference range of a bonus to give to the worker (“You typically consider offering a bonus between $0 and $50 for such work”). At this point in the scenario, we manipulated the worker’s use of AI: In the AI condition, participants read that “From your earlier conversation with the designer, you know that they used an AI system to create the ad,” whereas this sentence was absent in the No AI condition. Participants then answered the dependent measure of the study: “What amount would you give the designer as a bonus?” (followed by a 51-point slider scale from $0 to $50 in $1 increments with the slider button anchored at $0). Afterwards, participants answered exploratory measures and provided demographic information (age and gender) to complete the study. For Study 3 materials, see *SI*, Section 5.

**Study 4**

Study 4 tested the same hypotheses with the same design as in Study 3 but featured some changes in the scenario. Study 4 again tested (1) whether people would reduce compensation for workers using AI and (2) whether such a reduction in compensation would occur for both temporary and permanent workers.

As in Study 3, Study 4 employed a 2 (Temporary vs. Permanent Worker) × 2 (AI vs. No AI) between-subjects design. As in previous studies, participants across the four conditions ($N$ = 398) imagined a scenario in which they were running a small business. Unlike in previous studies, however, participants imagined “launch[ing] a new product” and working with a *web* designer to “create a dedicated landing page for it” (rather than creating a social media ad with a graphic designer). The descriptions for worker type were changed as well: Instead of a “salaried” graphic designer and a graphic designer “from an online freelancing platform,” the workers were respectively described as “full-time” and “freelance” web designers to maintain parallelism in wording. The quality of the workers’ outputs was controlled with a more specific description: “[Y]ou reviewed the landing page they designed, and you are impressed with the result. The page is visually appealing and functional.” We manipulated the use of AI by including or excluding the sentence, “During your earlier discussion, [the worker] mentioned that they used an AI tool to assist with the design.” Finally, participants chose a bonus for the worker on a different scale ($50 to $150, rather than $0 to $50). Aside from these differences, all other aspects of Study 4 mirrored those of Study 3. For Study 4 materials, see *SI*, Section 6.

**Study 5**

Studies 3 and 4 show that worker status (i.e., whether a worker is hired temporarily or permanently) does not moderate people’s tendency to reduce compensation when workers use AI. In Study 5, we tested whether having a prior history with the worker might attenuate such reduction in compensation.

We randomly assigned 200 participants to one of four conditions in a 2 (No History vs. Long History) × 2 (AI vs. No AI) between-subjects design. As in previous studies, participants imagined that they ran a small business, and the scenario details matched those of Study 2. In the No History condition, participants imagined hiring a new graphic designer with whom they had no prior history of working together (“Over the course of the year, you have hired different graphic designers… Today you found another graphic designer who agreed to create a new ad”). In the Long History condition, participants imagined working with one designer for a long time (“Over the course of the past 4 years, you have been working with a graphic designer… Today you and the designer discussed creating a new ad”). The worker’s use of AI was manipulated the same way as in Study 2, by including or excluding the sentence, “The graphic designer asked if they [could] work with an AI system for creating the ad, to which you agreed.” The study’s dependent measure asked participants to choose the amount of payment to offer the designer on a slider scale from $0 to $100. After choosing the payment amount, participants reported age and gender to complete the study. For Study 5 materials, see *SI*, Section 7.

**Study 6**

Study 6 further tested the robustness of the AI penalty by making changes in five main aspects. First, we used an alternative method to measure the dependent variable, replacing a slider scale with a ternary choice measure. Second, we examined the AI penalty in the context of productivity growth by emphasizing that workers’ productivity increased (with or without the use of AI)—a realistic workplace outcome not explored in previous studies. Third, we investigated a different category of workers, namely, part-time workers (rather than freelancers). Fourth, we modified the scenario to involve multiple workers rather than a single worker, which better reflects more common real-world business contexts. Lastly, we made the scenario more abstract by removing details on payment structures, specific tasks, and industry context.

We randomly assigned 401 participants to one of four conditions in a 2 (Temporary vs. Permanent Worker) × 2 (AI vs. No AI) between-subjects design. In all conditions, participants imagined that they ran a small business with five employees, but these employees were described either as “part-time employees” in the Temporary Worker condition or “full-time employees” in the Permanent Worker condition. More importantly, the scenario in this study described a growth in productivity: “At the end of the year, you notice that the productivity of your [full-time / part-time] employees has slightly increased compared to the previous year.” This productivity growth was attributed to the use of AI in the AI condition (“This increase in productivity was mainly due to the new AI tools you provided them earlier in the year”), or no such attribution was made in the No AI condition.

Aside from the information about productivity growth, the scenario did not feature any concrete details like those featured in previous studies (e.g., specific tasks like creating a social media ad or a landing page, or industry contexts like graphic or web design). Based on the abstract information alone, then, participants answered the dependent measure with three choices: “Assuming that the current level of productivity [driven by the AI tools (this phrase was inserted accompanied by commas only in the AI condition)] can be sustained next year, would you adjust the compensation for the [full-time / part-time] employees? That is, would you increase or decrease the compensation, or keep it the same? (Decrease the compensation / Keep the compensation the same / Increase the compensation).” After answering this question, participants answered an exploratory measure and provided demographic information to complete the study. For Study 6 materials, see *SI*, Section 8.

**Study 7**

Having found that people reduce compensation for workers using AI (Studies 1 and 2) and having confirmed that this effect was robust (Studies 3–6), we conducted Study 7 to examine whether this AI penalty might be explained by a perceived reduction in credit deservingness associated with the use of AI. To this end, we replicated the design of Study 1 and additionally assessed how much credit participants thought the workers deserved for their work.

We recruited 303 workers from Prolific and randomly assigned them to either the AI or No AI condition. As in Study 1, participants imagined hiring a graphic designer who created a social media ad for them either using an AI system or not (i.e., no indication of their AI use). After participants answered the dependent measure—their choice of the hypothetical payment amount for the designer—they answered three questions assessing *how much credit they thought the designer deserved for their work* (hereafter, *credit deservingness*): (1) “How much credit do you think the graphic designer deserves for creating the ad? (1 = *No credit at all*, 7 = *All the credit*)”; (2) “How responsible do you think the graphic designer was for creating the ad? (1 = *Not at all responsible*, 7 = *Completely responsible*)”; (3) “How important do you think the graphic designer’s role was in creating the ad? (1 = *Not at all important*, 7 = *Extremely important*).” Participants then reported age and gender to complete the study. For Study 7 materials, see *SI*, Section 9.

**Study 8**

We conducted Study 8 with two goals in mind: to test again the simple mediation model from Study 7 and to explore whether the mediation might be moderated by the extent to which the act of reducing worker compensation was *permissible*. We hypothesized that when it is less permissible to reduce worker compensation—for example, because their compensation is protected by an employment contract—then the AI penalty through credit deservingness might weaken. We thus adapted the scenario from Study 4 not only to manipulate whether a worker used AI, but also to create two different situations where reducing worker compensation was more or less permissible.

We randomly assigned 281 participants to one of four conditions in a 2 (AI vs. No AI) × 2 (More vs. Less Permissible [to Reduce Worker Compensation]) between-subjects design. All participants imagined that they ran "a small company that sells consumer products" and that they "decided to launch a new product and wanted to create a dedicated landing page for it." Unlike in Study 4, Study 8 scenario described working with an *in-house* web designer and did not explicitly describe them as "full-time" or "freelance." Moreover, the scenario explicitly described the task as an *extra* task (outside of their regular tasks) voluntarily accepted by the worker based on mutual agreement with the participant: "You asked your web designer whether they would be interested in creating the landing page. They agreed to do so as an extra task, in addition to their regular tasks." As in Study 4, Study 8 scenario controlled the quality of the work output to be high but explicitly mentioned a realistic amount of time taken to complete the task: "Within just three days of taking on the task, the designer delivered a landing page that was visually appealing and functioned exactly as you envisioned."

At this point, we manipulated the use of AI with a more realistic detail that was not included in Study 4. In the AI condition, participants read: "During your conversation about the task, the designer mentioned that they used AI tools (ChatGPT and Midjourney) to create the page." In the No AI condition, this sentence was omitted. Finally, participants imagined choosing the amount of bonus for the designer ("You are considering offering a bonus for this good work").

At this point in the scenario, we manipulated the extent to which it was permissible to reduce the worker's compensation. Specifically, in the Less Permissible condition, participants read "Your company writes into employment contracts to give bonus payments of $100 for extra tasks such as this," whereas in the More Permissible condition, they read, "Your company has in the past consistently given bonus payments of $100 for extra tasks such as this." Thus, we presented the same reference bonus amount of $100 in all conditions, but this amount reflected either a strict term in the employment contracts (in the Less Permissible condition) or the company's past behavior (in the More Permissible condition). By explicitly telling participants that the amount was written in the employment contract, we sought to convey that reducing the bonus amount would not be permissible, or at least be less permissible than simply deviating from the past behavior. After reading this last sentence of the scenario, participants chose the bonus amount for the designer on a 201-point slider scale from $0 to $200. Participants then answered the three mediator items measuring credit deservingness (the same items as in Study 7) and reported age and gender to complete the study. For Study 8 materials, see *SI*, Section 10.

**Study 9**

Studies 1–8 show that people penalize workers when the workers receive help from AI—but is this effect unique to AI? In other words, do people penalize workers for receiving help from AI specifically or for receiving help from any entity at all? Study 9 was conducted to explore these questions. We used the same scenario as in Study 2 and included the same No AI and AI conditions (respectively labeled “No Help” and “Help From AI” conditions). More importantly, however, we added a third condition (labeled “Help from Human” condition) whose scenario described a worker receiving help from *another human worker*. We reasoned that if people penalize workers for receiving help from any entity, we should observe a reduction in compensation in both the Help From AI and Help From Human conditions. However, if people penalize workers for receiving help specifically from AI, then we should observe a reduction in compensation only in the Help From AI condition and not in the Help From Human condition. Therefore, Study 9 allowed us to test whether our effect is indeed an “AI Penalization” effect, rather than an “Assistance Penalization” effect.

We randomly assigned 471 participants to one of three conditions (No Help vs. Help From AI vs. Help From Human) in a between-subjects design. As in Study 2, all participants entered Study 9 immediately after completing another study for an unrelated project. The No Help and Help From AI conditions were respectively identical to the No AI and AI conditions from Study 2. Participants in all conditions imagined running a small business and hiring a graphic designer to create a social media ad. In the Help From AI condition, participants learned that the designer received help from an AI system to create the ad (“The graphic designer asked if they can work with an AI system for creating the ad, to which you agreed”), whereas in the Help From Human condition, participants learned that the designer received help from another human to create the ad (“The graphic designer asked if they can work with another graphic designer for creating the ad, to which you agreed”). In the No Help condition, the sentence about the designer receiving help (from either AI or another human) was omitted. In all conditions, participants chose the payment for the designer (i.e., answered the question, “How much payment would you offer them?” on a 101-point slider scale from \$0 to \$100). Participants then reported demographic information to complete the study. For Study 9 materials, see *SI*, Section 11.

**Study 10**

So far, we have used scenarios to investigate whether and why people might reduce compensation for workers using AI. In Study 10, we examined whether this reduction in compensation extends to real payments made to real workers. To do so, we first recruited participants to work on a task either using AI or not. We then recruited a second group of participants to take on the role of manager that allocates real monetary bonuses among some of the previous workers with the knowledge of which workers had used AI. We thus tested whether the AI penalty would be replicated with real monetary compensation to real workers.

In addition, by recruiting different participants to work on the same task, Study 10 allowed the quality of workers' outputs to vary naturally. Study 10 then examined the effect of AI use on worker compensation while holding constant this natural variation in workers' output quality. In some of the previous studies, we artificially held workers' output quality constant through scenario details (e.g., by telling participants that the outputs of the workers who used AI and those who did not use AI were both "impress[ive]," "good," or "appealing" as in Studies 4 and 8). In other studies, we allowed participants to make different inferences about workers' (future) output quality in the AI and No AI conditions by not providing such details in the scenario (e.g., Studies 1 and 5). However, the scenarios used in these studies may have resulted in a perceived difference in workers' output quality between the AI and No AI conditions. For example, a "good work" done with AI and a "good work" done without AI may not be of the same quality in participants' minds. So, it is possible that such a difference in perceived quality of workers' outputs between the AI and No AI condition could drive the difference in worker compensation. Study 10 thus addressed this concern by allowing a natural variation in workers' output quality and holding this variation constant.

Study 10 was conducted in two parts. In Part 1, we recruited from Prolific 60 participants (hereafter, *the workers*) to work on a task as gig workers. The workers were first thanked for participating in the study and were informed that they would "write a short social media post to promote a fictional product." They were further told that a judge or judges would evaluate the social media posts in terms of "how likely they [were] to encourage readers to learn more about the product," and that if their post ranked among the top 50% of all the posts, they would receive a bonus payment of $0.30. (We paid out these bonuses shortly after the data collection in Part 1.) We then asked the workers, "Are you motivated to write a social media post that earns a bonus? [Yes, I am motivated / No, I am not really motivated]." We used this item to screen out submissions from unmotivated workers when presenting their work to managers.

The workers then received the instructions for the task. Specifically, they were asked to imagine that they were a social media manager for a product and to "write a 2-4 sentence social media post that grabs attention and encourages readers to learn more about the product." They then received information on the product name ("Beannovation") and product description ("Beannovation is a cutting-edge coffee maker… Brew a rich, barista-quality cup in just 3 minutes…"). The workers received more detailed instructions for writing the social media post ("Highlight the key benefit(s)… End with a call to action") and saw an example of a social media post. The exact materials presented to the workers are included in *SI*, Section 12, "Part 1."

At this point, we manipulated the workers' use of AI. Below the aforementioned example of a social media post, a randomly selected half of the workers were presented with the instructions to use an AI tool for the task: "For this task, please use our ChatGPT interface by clicking the button below." Below these instructions was a button that the workers could click, and once they clicked the button, a new browser tab opened and presented the workers with an interface that allowed them to directly interact with ChatGPT to receive assistance for the task. This AI assistance feature was added using the free program *G4R ("GPT for Researchers")* [34]. After the workers used ChatGPT to draft or refine their social media post, they returned to the original study page and entered their social media post in a text box. For the other half of the workers, no AI tool was provided, and these workers wrote a social media post on their own. (In this No AI condition, submissions from workers who later reported using AI were excluded for managers' evaluation.) After writing their social media posts, all workers answered exploratory questions which asked whether and how they used an AI tool for the task (either within or outside the study) and provided demographic information to complete the study; see *SI*, Section 12, "Part 1," for all measures used in the study.

From the 60 social media posts submitted in Part 1, we selected stimuli to be used in Part 2. Specifically, we selected four social media posts that were of similar length and quality (as judged by the first author), two of which were produced by workers using AI and the other two produced by workers not using AI. These four submissions were chosen as workers' output to be evaluated by managers in Part 2.

In Part 2, we recruited from Prolific 140 participants (hereafter, *the managers*) who would take on the role of managers overseeing the workers from Part 1. The managers were first thanked for participating in the study and were informed that they would "make decisions as a manager overseeing four gig workers who participated in our previous study on Prolific." They were informed that the workers had previously written social media posts for a product and were presented with the same information on product name and description that had been presented to the workers.

Each of the 140 managers were asked to "review the submissions from four workers and allocate a total bonus of $1.00 among them." These submissions from four workers referred to those social media posts selected as stimuli for Part 2. The managers were further informed that the bonus amount they assign to each worker would "serve as input to determine the actual bonus payments" that the workers receive. Before reviewing the social media posts, the managers received a note in red font which stated that "[s]ome workers were provided with an AI tool to assist in creating their social media posts" and that whether a given worker used AI would be indicated for each post.

Next, each of the 140 managers reviewed the same four social media posts. Importantly, these posts came with labels that indicated whether the workers used AI; see *SI*, Section 12, "Part 2." For example, above the first social media post was the label "Submission by Worker 1 (who used an AI tool)" with the parenthetical remark in red font. We randomized the order of the four social media posts, as well as which two of the posts would be labeled as assisted by AI ("who used an AI tool") and which two would be labeled as unassisted by AI ("who did not use an AI tool"). Thus, the labels indicating whether the worker used AI or not were affixed to the posts *independently* of whether the workers actually used AI or not. After reviewing the four social media posts by four workers, each manager decided the amount of bonus to give to each worker by entering a number between 0 and 100 (representing $1.00 or 100 cents) in each of four text boxes ("Bonus for Worker 1: ___; Bonus for Worker 2: ___; [and so on]"). (We later used managers' bonus decisions to calculate and pay the average bonus amount to each of the four workers within a few days of data collection.) The managers then provided demographic information to complete the study. For Study 10 materials, see *SI*, Section 12.

**Study 11**

We recruited 505 participants (hereafter, *the managers*) and randomly assigned them to one of two conditions (No AI vs. AI) in a between-subjects design. All the managers were told that they would make a decision as a "manager overseeing a gig worker who participated in our previous study." The managers then received the same information as in the Part 2 of Study 10: that the workers had previously crafted social media posts about a product, along with the product name and description presented to the workers. Next, the managers were informed that they would act as a manager overseeing *one of the workers* and were asked to review the submission from their worker and choose a bonus amount between $0.00 and $5.00 for their worker. They further learned that the bonus amount they choose will "serve as input to determine the actual bonus payment" the worker receives.

At this point, we manipulated the worker's purported use of AI by inserting a note indicating the worker's use of AI in the AI condition ("Note: This worker used an AI tool to create their social media post") or not inserting this note in the No AI condition. Whether this note was included or not was random and independent of whether the focal worker actually used AI. The managers then saw the social media post created by their worker. Here, we randomly selected and presented one of the four social media posts used in Part 2 of Study 10 as their worker's submission. After reading their worker's social media post, the managers chose an amount of bonus for their worker using a slider scale ($0.00 to $5.00 in $0.01 increments). (We later used the managers' bonus decisions to calculate and pay the average bonus amount to each of the four workers within a few days of data collection, in addition to previous sets of bonuses

we had paid them). The managers then answered an unrelated question for a different research project and reported demographic information to complete the study. For Study 11 materials, see *SI*, Section 13.

**Study 12**

Study 12 delved deeper into the mechanism underlying the AI penalty by experimentally manipulating two worker inputs expected to be affected by AI use: effort and agency. Specifically, we investigated whether reductions in compensation for AI-assisted workers arise because evaluators perceive them as exerting less effort, relinquishing agency over core aspects of their work, or both.

We recruited 809 participants in two batches and randomly assigned them to one of eight conditions in a 2 (AI vs. No AI) × 2 (Agency: Low vs. High) × 2 (Effort: Low vs. High) between-subjects design. All participants imagined a modified version of the "graphic designer" scenario used in previous studies (i.e., running a small business and hiring a graphic designer to create a social media ad for the business). Unlike in previous studies, we more tightly controlled the quality of the worker's output by telling participants, "To evaluate the ad, you showed it to four of your regular customers, who rated its quality at an average of 8.5 out of 10, comparable to other ads you've recently used." We also manipulated AI use differently than in previous studies, by telling participants that the worker used either "an AI tool" (AI condition) or "standard design software" (No AI condition) to create the ad.

Next, we manipulated the perceived agency of the worker. In the low agency condition, participants imagined that the worker's tool—rather than the worker—largely determined the work output ("The AI tool / design software provided ready-made ad templates, and the designer selected one to produce the final design"). In the high agency condition, participants imagined that the designer exercised substantial control over the creative process and used the tool primarily for refinement ("The designer provided a highly specific prompt… to guide the AI toward the specific design they had in mind" [AI condition] or "The designer generated the ad's core concept and messaging on their own… performed limited manual polishing work at the end using the design software" [No AI condition]). We then manipulated the perceived effort of the worker by telling participants either that "The designer spent about 30 minutes on the ad and described it as a quick, minimal-effort job, involving few iterations" (low effort condition) or that "The designer spent about about 4 hours on the ad and described it as a careful, high-effort job involving multiple iterations" (high effort condition).

All participants then chose a bonus for the worker using a slider scale ($0 to $50 in $1 increments). They then answered a series of four questions in a random order: two questions measuring the worker's perceived effort (e.g., "How much effort do you think the designer put

into creating the ad?" [1 = *Very little*, 7 = *Very much*]) and two questions measuring the worker's perceived agency (e.g., "To what extent do you think the ad would reflect the designer's own work rather than the tool used?" [1 = *Not at all*, 7 = *Very much*]). Finally, participants reported perceived originality of the work ("How original do you think the ad was?" [1 = *Not original at all*, 7 = *Very original*]) and the extent to which they thought the worker deserved credit for the work ("How much credit do you think the graphic designer deserves for creating the ad?" [1 = *No credit at all*, 7 = *All the credit*]). For Study 12 materials, see *SI*, Section 14.

**Study 13**

Study 13 examined whether the AI penalty observed across diverse creative tasks (e.g., creating social media ads, designing a web page, and writing social media posts) is specific only to creative work (where creative agency would be salient), or whether it also emerges in routine types of work.

We randomly assigned 301 participants to either the AI or No AI condition. In each condition, participants imagined a scenario from Study 12 that was modified to feature a routine task. Specifically, all participants imagined running a small business but hiring "an analyst to compile and summarize quarterly sales data into a standardized report" for the business. We manipulated the worker's use of AI by either indicating that the analyst used an AI tool to produce the report or omitting any information about AI use. All participants then chose a bonus for the worker using the same slider scale as in Study 12. Afterwards, participants answered two questions measuring perceived effort (e.g., "How much effort do you think the analyst put into producing the report?" [1 = *Very little*, 7 = *Very much*]) and three questions measuring perceived agency (e.g., "To what extent do you think the report would reflect the analyst's own work?" [1 = *Not at all*, 7 = *Very much*]), presented in random order. Lastly, participants reported perceived originality of the work ("How original do you think the report was?" [1 = *Not original at all*, 7 = *Very original*]). For Study 13 materials, see *SI*, Section 15.

**References**

1. Maslej, N. *et al.* Artificial Intelligence Index Report 2024. (2024).
2. Paradis, E. *et al.* How much does AI impact development speed? An enterprise-based randomized controlled trial. *arXiv preprint arXiv:2410.12944* (2024).
3. Kanazawa, K., Kawaguchi, D., Shigeoka, H. & Watanabe, Y. *Ai, Skill, and Productivity: The Case of Taxi Drivers*. (2022).
4. Zhou, E. & Lee, D. Generative artificial intelligence, human creativity, and art. *PNAS nexus* **3**, pgae052 (2024).
5. Lin, L. *About 1 in 5 U.S. Workers Now Use AI in Their Job, up since Last Year*. https://www.pewresearch.org/short-reads/2025/10/06/about-1-in-5-us-workers-now-use-ai-in-their-job-up-since-last-year/ (2025).
6. Longoni, C., Fradkin, A., Cian, L. & Pennycook, G. News from generative artificial intelligence is believed less. in *Proceedings of the 2022 ACM Conference on Fairness, Accountability, and Transparency* 97–106 (2022).
7. Magni, F., Park, J. & Chao, M. M. Humans as creativity gatekeepers: Are we biased against AI creativity? *Journal of Business and Psychology* **39**, 643–656 (2024).
8. Ragot, M., Martin, N. & Cojean, S. Ai-generated vs. human artworks. a perception bias towards artificial intelligence? in *Extended abstracts of the 2020 CHI conference on human factors in computing systems* 1–10 (2020).
9. Reif, J. A., Larrick, R. P. & Soll, J. B. Evidence of a social evaluation penalty for using AI. *Proceedings of the National Academy of Sciences* **122**, e2426766122 (2025).
10. Kim, H. & Koo, T. K. The Impact of Generative AI on Syllabus Design and Learning. *Journal of Marketing Education* 02734753241299024 (2024).
11. Microsoft. *2024 Work Trend Index Annual Report*. 7 https://assets-c4akfrf5b4d3f4b7.z01.azurefd.net/assets/2024/05/2024_Work_Trend_Index_Annual_Report_6_7_24_666b2e2fafceb.pdf (2024).
12. Slack. The Workforce Index. (2024).
13. Leib, M., Köbis, N. & Soraperra, I. Does AI and human advice mitigate punishment for selfish behavior? An experiment on AI ethics from a psychological perspective. *Computers in Human Behavior* **171**, 108709 (2025).
14. Bindley, K. & Blunt, K. Tech Firms Aren't Just Encouraging Their Workers to Use AI. They're Enforcing It. *The Wall Street Journal* (2026).
15. Acemoglu, D. & Restrepo, P. Artificial intelligence, automation, and work. in *The economics of artificial intelligence: An agenda* 197–236 (University of Chicago Press, 2019).

16. Engberg, E., Koch, M., Lodefalk, M. & Schroeder, S. Artificial intelligence, tasks, skills and wages: Worker-level evidence from Germany. Preprint at https://www.econstor.eu/bitstream/10419/298567/1/1877476331.pdf (2023).
17. Felten, E., Raj, M. & Seamans, R. How will language modelers like chatgpt affect occupations and industries? *arXiv preprint arXiv:2303.01157* (2023).
18. Fossen, F. M., Samaan, D. & Sorgner, A. How are patented AI, software and robot technologies related to wage changes in the United States? *Frontiers in Artificial Intelligence* **5**, 869282 (2022).
19. Copestake, A., Marczinek, M., Pople, A. & Stapleton, K. AI and Services-Led Growth: Evidence from Indian Job Adverts. Preprint at https://copestake.info/workingpaper/akai/AKAI.pdf (2024).
20. Hui, X., Reshef, O. & Zhou, L. The short-term effects of generative artificial intelligence on employment: Evidence from an online labor market. *Organization Science* (2024).
21. Qiao, D., Rui, H. & Xiong, Q. AI and Jobs: Has the Inflection Point Arrived? Evidence from an Online Labor Platform. *arXiv preprint arXiv:2312.04180* (2023).
22. Leventhal, G. S. What should be done with equity theory? New approaches to the study of fairness in social relationships. in *Social exchange: Advances in theory and research* 27–55 (Springer, 1980).
23. Kahneman, D., Knetsch, J. L. & Thaler, R. Fairness as a constraint on profit seeking: Entitlements in the market. *The American economic review* 728–741 (1986).
24. Glikson, E. & Woolley, A. W. Human trust in artificial intelligence: Review of empirical research. *Academy of management annals* **14**, 627–660 (2020).
25. Adams, J. S. Towards an understanding of inequity. *The Journal of Abnormal and Social Psychology* **67**, 422–436 (1963).
26. Carrell, M. R. & Dittrich, J. E. Equity theory: The recent literature, methodological considerations, and new directions. *Academy of management review* **3**, 202–210 (1978).
27. Brynjolfsson, E. & McAfee, A. *The Second Machine Age: Work Progress and Prosperity in a Time of Brilliant Technologies*. (WW Norton & company, 2014).
28. Vanneste, B. S. & Puranam, P. Artificial intelligence, trust, and perceptions of agency. *Academy of Management Review* amr-2022 (2024).
29. Hermann, E., Puntoni, S. & Morewedge, C. K. GenAI and the psychology of work. *Trends in Cognitive Sciences* (2025).
30. Bankins, S. & Formosa, P. The ethical implications of artificial intelligence (AI) for meaningful work. *Journal of Business Ethics* **185**, 725–740 (2023).

31. Schweitzer, S. & De Cremer, D. When being managed by technology: does algorithmic management affect perceptions of workers' creative capacities? *Academy of Management Discoveries* **10**, 375–392 (2024).
32. *Temporary Jobs Are Growing Fast, But Temp Workers Have Few Legal Protections*. https://www.nelp.org/temporary-jobs-growing-fast-temp-workers-legal-protections/ (2019).
33. Jago, A. S., Raveendhran, R., Fast, N. & Gratch, J. Algorithmic management diminishes status: An unintended consequence of using machines to perform social roles. *Journal of Experimental Social Psychology* **110**, 104553 (2024).
34. Kim, J. How to Capture and Study Conversations Between Research Participants and ChatGPT: GPT for Researchers (g4r.org). Preprint at https://doi.org/10.48550/arXiv.2503.18303 (2025).